\documentclass[prd, preprintnumbers, amsmath, amssymb, nofootinbib, aps, superscriptaddress, letterpaper]{revtex4}

\usepackage{bm}
\usepackage{amsfonts}
\usepackage{mathrsfs}
\usepackage{graphicx}
\usepackage{amsmath}
\usepackage{float}
\usepackage{braket}
\usepackage{natbib}
\newcommand{\beginsupplement}{%
        \setcounter{table}{0}
        \renewcommand{\thetable}{S\arabic{table}}%
        \setcounter{figure}{0}
        \renewcommand{\thefigure}{S\arabic{figure}}%
     }

\begin{document}

\newcommand{\sgn}{\operatorname{sgn}}
\newcommand{\hhat}[1]{\hat {\hat{#1}}}
\newcommand{\pslash}[1]{#1\llap{\sl/}}
\newcommand{\kslash}[1]{\rlap{\sl/}#1}
\newcommand{\lab}[1]{}
\newcommand{\iref}[2]{}
\newcommand{\sto}[1]{\begin{center} \textit{#1} \end{center}}
\newcommand{\rf}[1]{{\color{blue}[\textit{#1}]}}
\newcommand{\eml}[1]{#1}
\newcommand{\el}[1]{\label{#1}}
\newcommand{\er}[1]{Eq.\eqref{#1}}
\newcommand{\df}[1]{\textbf{#1}}
\newcommand{\mdf}[1]{\pmb{#1}}
\newcommand{\ft}[1]{\footnote{#1}}
\newcommand{\n}[1]{$#1$}
\newcommand{\fals}[1]{$^\times$ #1}
\newcommand{\new}{{\color{red}$^{NEW}$ }}
\newcommand{\ci}[1]{}
\newcommand{\de}[1]{{\color{green}\underline{#1}}}
\newcommand{\ke}{\rangle}
\newcommand{\br}{\langle}
\newcommand{\lb}{\left(}
\newcommand{\rb}{\right)}
\newcommand{\lbk}{\left[}
\newcommand{\rbk}{\right]}
\newcommand{\blb}{\Big(}
\newcommand{\brb}{\Big)}
\newcommand{\nn}{\nonumber \\}
\newcommand{\p}{\partial}
\newcommand{\pd}[1]{\frac {\partial} {\partial #1}}
\newcommand{\cd}{\nabla}
\newcommand{\cc}{$>$}
\newcommand{\bqa}{\begin{eqnarray}}
\newcommand{\eqa}{\end{eqnarray}}
\newcommand{\bqe}{\begin{equation}}
\newcommand{\eqe}{\end{equation}}
\newcommand{\bay}[1]{\left(\begin{array}{#1}}
\newcommand{\eay}{\end{array}\right)}
\newcommand{\eg}{\textit{e.g.} }
\newcommand{\ie}{\textit{i.e.}, }
\newcommand{\iv}[1]{{#1}^{-1}}
\newcommand{\st}[1]{|#1\ke}
\newcommand{\at}[1]{{\Big|}_{#1}}
\newcommand{\zt}[1]{\texttt{#1}}
\newcommand{\non}{\nonumber}
\newcommand{\m}{\mu}
\def\xa{{m}}
\def\xA{{m}}
\def\xb{{\beta}}
\def\xB{{\Beta}}
\def\xd{{\delta}}
\def\xD{{\Delta}}
\def\xe{{\epsilon}}
\def\xE{{\Epsilon}}
\def\xve{{\varepsilon}}
\def\xg{{\gamma}}
\def\xG{{\Gamma}}
\def\xk{{\kappa}}
\def\xK{{\Kappa}}
\def\xl{{\lambda}}
\def\xL{{\Lambda}}
\def\xo{{\omega}}
\def\xO{{\Omega}}
\def\xvp{{\varphi}}
\def\xs{{\sigma}}
\def\xS{{\Sigma}}
\def\xt{{\theta}}
\def\xvt{{\vartheta}}
\def\xT{{\Theta}}
\def \Tr {{\rm Tr}}
\def\CA{{\cal A}}
\def\CC{{\cal C}}
\def\CD{{\cal D}}
\def\CE{{\cal E}}
\def\CF{{\cal F}}
\def\CH{{\cal H}}
\def\CJ{{\cal J}}
\def\CK{{\cal K}}
\def\CL{{\cal L}}
\def\CM{{\cal M}}
\def\CN{{\cal N}}
\def\CO{{\cal O}}
\def\CP{{\cal P}}
\def\CQ{{\cal Q}}
\def\CR{{\cal R}}
\def\CS{{\cal S}}
\def\CT{{\cal T}}
\def\CV{{\cal V}}
\def\CW{{\cal W}}
\def\CY{{\cal Y}}
\def\BC{\mathbb{C}}
\def\BR{\mathbb{R}}
\def\BZ{\mathbb{Z}}
\def\sA{\mathscr{A}}
\def\sB{\mathscr{B}}
\def\sF{\mathscr{F}}
\def\sG{\mathscr{G}}
\def\sH{\mathscr{H}}
\def\sJ{\mathscr{J}}
\def\sL{\mathscr{L}}
\def\sM{\mathscr{M}}
\def\sN{\mathscr{N}}
\def\sO{\mathscr{O}}
\def\sP{\mathscr{P}}
\def\sR{\mathscr{R}}
\def\sQ{\mathscr{Q}}
\def\sS{\mathscr{S}}
\def\sX{\mathscr{X}}

\def\slz{SL(2,Z)}
\def\slr{$SL(2,R)\times SL(2,R)$ }
\def\ads{${AdS}_5\times {S}^5$ }
\def\adst{${AdS}_3$ }
\def\sun{SU(N)}
\def\ad#1#2{{\frac \delta {\delta\sigma^{#1}} (#2)}}
\def\bqf{\bar Q_{\bar f}}
\def\nf{N_f}
\def\sunf{SU(N_f)}

\def\dcirc{{^\circ_\circ}}

\author{Morgan H. Lynch}
\email{morgan.lynch@technion.ac.il}
\affiliation{Department of Electrical Engineering,
Technion - Israel Institute of Technology, Haifa 32000, Israel}
\author{Eliahu Cohen}
\email{eliahu.cohen@biu.ac.il}
\affiliation{Physics Department, Centre for Research in Photonics, University of Ottawa, Advanced Research Complex,
25 Templeton, Ottawa, Ontario, Canada K1N 6N5}
\affiliation{Faculty of Engineering and the Institute of Nanotechnology and Advanced Materials,
Bar Ilan University, Ramat Gan 5290002, Israel}
\author{Yaron Hadad}
\email{yaronhadad@gmail.com}
\affiliation{Department of Mathematics, University of Arizona, Tucson, AZ, 85721, USA}
\author{Ido Kaminer}
\email{kaminer@technion.ac.il}
\affiliation{Department of Electrical Engineering,
Technion - Israel Institute of Technology, Haifa 32000, Israel}

\title{Accelerated-Cherenkov radiation and signatures of radiation reaction}
\date{\today}

\begin{abstract}
In this manuscript we examine an accelerated charged particle moving through an optical medium, and explore the emission of accelerated-Cherenkov radiation. The particle's reaction to acceleration creates a low-frequency spectral cutoff in the Cherenkov emission that has a sharp resonance at the superluminal threshold. Moreover, the effect of recoil on the radiation is incorporated kinematically through the use of an Unruh-DeWitt detector by setting an energy gap, i.e., the change in electron energy, to the recoil energy of the emitted photon. The simultaneous presence of recoil and acceleration conspire to produce a localized resonance peak in the emission. These theoretical considerations could be used to construct high precision tests of radiation reaction using Cherenkov emission under acceleration.
\end{abstract}


\maketitle

\section{Introduction}

The Unruh-DeWitt detector \cite{unruh1, dewitt} has found a variety of applications across several fields of physics. Its most famous manifestation, and original use, was in the discovery of the Unruh effect \cite{unruh1}. There, the transition rate of a uniformly accelerated Unruh-DeWitt detector, with proper acceleration $\bar{a}$, implied that Minkowski vacuum was comprised of a thermalized bath of particles at temperature $T = \frac{\bar{a}}{2\pi}$. In addition to its original use in exploring the particle content of general relativistic space times, it has also played a key role in the exploration of relativistic quantum information \cite{louko, bruschi, hu, bruschi1, lee}. This wide ranging applicability is born out of the fact that it enables an exploration of particle systems independent of the major details, i.e. it only cares about the change in energy. This has the effect of taking an otherwise complicated process and rendering it a two level system like an atom with one excited state or a ``qubit" of information. By ignoring the fine details and focusing on the particle's quantized recoil due to its radiation, the Unruh-Dewitt detector approach may be able to provide a new perspective on a long standing question: the problem of radiation reaction \cite{dirac}. Put simply, radiation reaction is the effect that photon emission has on a charged particle's trajectory while it radiates, e.g. a recoil correction due to the emitted photon's momentum. This is particularly exciting since recent theoretical and experimental works involving high intensity lasers and channeling radiation \cite{thomas, cole, poder, wistisen, chen} have reported the first signatures of radiation reaction which appears to have quantum aspects involved in them \cite{dipiazza, yaron, burton, iiderton}. 

Here we develop the Unruh-DeWitt detector formalism for its use in incorporating recoil and apply it to Cherenkov radiation \cite{cherenkov}. Moreover, we also include acceleration in the analysis to gain insight into the effects of radiation reaction beyond the conventional models that consider the recoil as a result of Larmor radiation \cite{jackson}. Fundamentally, this method is based on quantum field theory in curved spacetime for analyzing radiation emission \cite{davies, winitzki}. By incorporating the techniques used to explore radiation in more exotic spacetimes, we can gain considerable insight into the simpler setting of Minkowski space. To explore vector currents coupled to photons via the QED interaction, we use spacetime trajectories/world lines that are often utilized in quantum field theory in curved spacetime to investigate the emission of radiation \cite{parker, hawking, unruh1, muller, matsas1, matsas2, matsas3, lynch, lynch1} and apply them to superluminal velocities with acceleration in an optical medium, revealing a novel accelerated-Cherenkov effect. We derived a generalized Frank-Tamm  \cite{cherenkov, frank, ido} formula that simultaneously takes into account the quantum recoil due to a single photon emission as well as the acceleration; the inclusion of both effects yields a novel resonant spectral cutoff at the superluminal threshold. We also demonstrate how to incorporate the recoil, e.g. radiation reaction, by using the recoil momenta of the emitted photon as the change in energy of the electron. The preliminary example of Cherenkov radiation is given to outline the functionality of the technique, how to incorporate the recoil, and is then applied to the case where an electron is accelerating in the superluminal regime. The presence of the acceleration sets an energy scale that sharply cuts off the photon emission. This energy scale is quite tunable due to the presence of a strong resonance as the electron velocity approaches the in-medium speed of light. Starting from the cutoff energy, the emitted photon spectra rapidly climbs and asymptotes to the flat Cherenkov emission spectra. The incorporation of the particle recoil due to the emission of a single photon quanta affects the higher frequencies to produce a monotonically decreasing spectra. The combination of both yields a peak in the Cherenkov emission spectrum which may be measurable experimentally \cite{kimura, yhu, henstridge}. 

Section 2 of this paper explores photon emission in QED \cite{peskin} and its relationship to an Unruh-DeWitt detector with the only assumption being that the electron momentum is much bigger than the emitted photon momentum (i.e. the weak recoil approximation). By propagating the detector along an inertial, i.e. non accelerated, world line moving superluminally through an optical medium, we are then able to analyze Cherenkov emission. The overall computational technique is presented and we show how to take into account the quantum recoil of photon emission by setting the Unruh-DeWitt detector energy to the recoil energy produced by the photon emission. Then, in Sec. III we retrace the computation except we move the detector along a uniformly accelerated world line. The effects of incorporating acceleration, including the particle recoil due to the interaction with the second-quantized photon field, is then explored. Strong signatures of the acceleration can be tuned by a resonance at the superluminal threshold and its combined effect with recoil produces a broad peak in the emission spectra. Finally, a summary of conclusions are presented and we have included all pertinent derivations in the appendix. All calculations are performed using the natural units $\hbar = c = k_{B} =1$.  

\section{Cherenkov Emission via Unruh-DeWitt detectors}

The emission of radiation produced by inertial currents offers the easiest setup to analyze recoil since the kinematics are purely Minkowskian. This setting provides a rather simple introduction to both radiation emission as well as the recoil. We begin by examining the emission of a photon in a refractive medium, $\hat{A}^{\m}(x)$, using the current interaction for QED \cite{peskin}, $\hat{S}_{I} = \int d^{4}x \hat{j}_{\m}(x)\hat{A}^{\m}(x)$. To model the electron current semi-classically we will make use the charged current coupled to an Unruh-DeWitt detector, $\hat{j}_{\m}(x) = u_{\m}\hat{q}(\tau)\delta^{3}(x-x_{tr})$. Here we defined the Heisenberg evolved charge monopole moment operator $\hat{q} = e^{i\hat{H}\tau}\hat{q}(0)e^{-i\hat{H} \tau}$ where $\hat{q}(0)$ is defined as $\hat{q}(0)\ket{E_{i}} = \ket{E_{f}}$ with $E_{i}$ and $E_{f}$ the initial energy and final energy of the electron moving along the trajectory, $x_{tr}$, of the current with four velocity $u_{\m}$. The energy gap of the detector is $\Delta E = E_{f} -E_{i}$ and the charge $q$ of the electron is given by $q =\vert \bra{E_{f}} \hat{q}(0)\ket{E_{i}} \vert$ \cite{davies, unruh1, matsas1, matsas2, matsas3}.

\begin{figure}[H]
\centering  
\includegraphics[scale=.50]{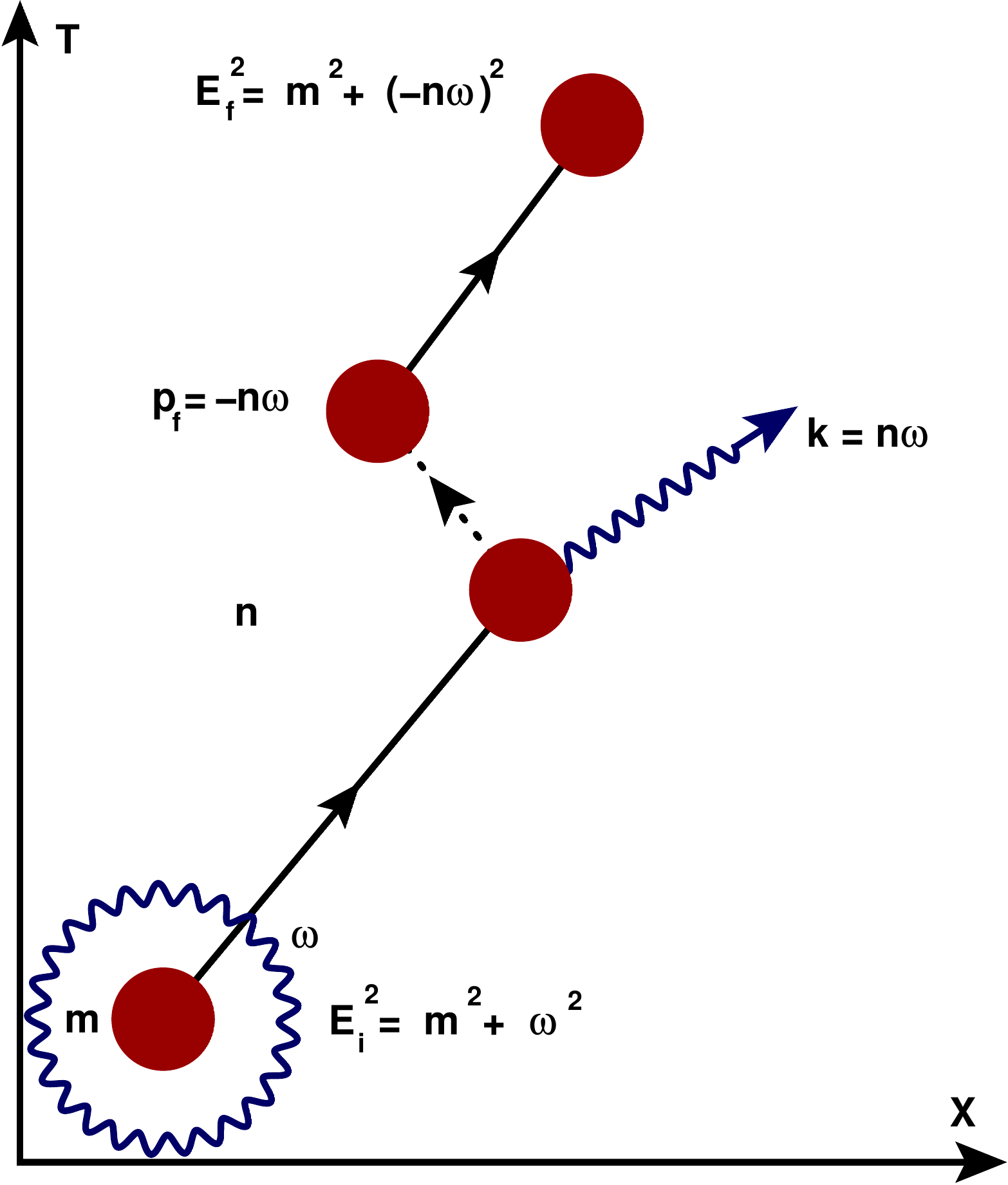}
\caption{The conceptual foundation for an initial dressed electron subject to recoil after photon emission. This assumption is used to define the energy gap of the Unruh-DeWitt detector. The photon energy contributes to the electrons mass and the non trivial dispersion produced by the indexed medium produces the recoil momentum when emitted. The kinetic energy produced by this recoil defines our energy gap Eq. (\ref{recoil}). This phenomenon can also be viewed as a consequence of mass renormalization in an optical medium.}	
\label{photon}
\end{figure}

In the case of superluminal motion through an indexed medium, the emission of radiation accompanied by the transition of an internal energy level is associated with the anomalous Doppler effect \cite{ginsburg}. This extra degree of freedom will allow us to incorporate electronic excitations to higher energy, i.e. the recoil. With the intent to examine Cherenkov radiation, i.e. photon emission from a charge moving faster than the speed of light in medium, we formulate the following amplitude; $\mathcal{A} = i\bra{\mathbf{k}}\otimes \bra{E_{f}}\hat{S}_{I}\ket{E_{i}}\otimes \ket{0}.$ The square of the this quantity determines the transition probability per momentum of the final state photon. Note, by having a detector we are relieved of integrating over the final electron momentum states. Constructing the response function yields
\bqa
\Gamma = q^{2} \int d\xi e^{-i\Delta E \xi/\gamma}  V_{\m \nu}[x',x]G^{\m \nu}[x',x]. \label{response}
\eqa
Here we are integrating over the laboratory time $\xi$. The ``velocity tensor" $V_{\m \nu} = v_{\m}v_{\nu}$ determines the trajectory of the electron with $v^{\m} = (1, 0, 0, \beta)$ given by the lab frame velocity. Computation of the emission spectrum yields the Frank-Tamm formula \cite{frank}, see the appendix for details. Hence,
\bqe
\frac{d \Gamma}{d \omega} = \alpha \beta  \lbk  1-\cos^{2}{(\theta_{c})} \rbk.\label{ft}
\eqe
We note the prefactor is nothing more than the fine structure constant $\alpha = \frac{q^{2}}{4 \pi}$. This well know expression characterizes the emission of Cherenkov photons by charged particles in the superluminal regime. Note that it scales as $\beta$ and is independent of the frequency modulo the frequency dependence in the index of refraction. The Cherenkov emission angle also contains a contribution from the energy gap which can be used to explore the anomalous Doppler effect and other phenomena  \cite{ginsburg,xihang}, $\cos{(\theta_{c})} = \frac{1}{n\beta}+\frac{\Delta E}{n\beta \omega \gamma}$. The energy gap serves to deform the emission angle making it smaller if the energy gap is positive (excitations) or larger if the energy gap is negative (decays). If we are examining the recoil from a photon emission we can also determine the explicit form for the energy gap. Considering an initial ``dressed" electron with total energy comprised of the electron rest mass and the emitted photons energy $E_{i} = \sqrt{m^2+\omega^2}$ and a final state electron energy comprised of a bare electron and recoil momentum $k=n\omega$ (where n is the frequency-dependent index of refraction) produced by the photon emission, $E_{f} = \sqrt{m^2+(n\omega)^2}$, see Fig. \ref{photon} for an illustrative example. The recoil momentum, $n\omega$, is characteristic of the non-trivial dispersion relation of the photon. As such, we can determine the energy difference to be
\bqa
\Delta E  =   \frac{(n^{2}-1)\omega^{2}}{2m}.
\label{recoil}
\eqa
Here, the energy gap of the Unruh-DeWitt detector is defined as the difference between the energies of the electron before and after emission. This specific energy gap, Eq. (\ref{recoil}), is present whether the photon is emitted at constant velocity in the superluminal regime or by acceleration. The energy gap is a by-product of mass renormalization where the electron's energy can be enhanced on account of it propagating in a material with dispersion \cite{tsytovich} and exists independent of whether or not the charged particle is accelerated. In general when in the presence of an optical medium one can always define the energy gap in such a way. There may, in fact, even be other terms which characterize other processes as well but we leave the examination of those to a later work. From this relation we also obtain the fully relativistic quantum corrected Cherenkov emission angle \cite{ginsburg}. Our Cherenkov angle, including the recoil correction, is then given by
\bqe
\cos{\theta_{c}} = \frac{1}{n\beta}+\frac{(n^2-1)\omega}{2m\gamma n\beta}.
\label{doppler}
\eqe
As such, by defining the energy gap in accordance with mass renormalization, we have derived the Frank-Tamm formula generalized to the exact quantum result that includes the reaction of the particle to its recoil by the photon emission \cite{sokolov,ido}. Hence, 

\bqe
\frac{d \Gamma}{d \omega} = \alpha \beta  \lbk  1-\frac{1}{n^{2}\beta^{2}} \lb  1+\frac{(n^{2}-1)\omega}{m\gamma} + \frac{(n^{2}-1)^{2}\omega^{2}}{4m^2\gamma^{2}} \rb \rbk .
\eqe
To conclude this section, we have outlined the general procedure for the computation of the emission rate of Cherenkov radiation, including recoil, via the use of an Unruh-DeWitt detector. We outlined the general idea for how to incorporate recoil with the Unruh-DeWitt detector; this procedure is discussed in more detail in the appendix. What is most important is the fact that we are able to successfully incorporate recoil into the emission process. In the following section we will follow the same procedure while placing the current along an accelerated trajectory which is also superluminal. We then explore the emission spectrum of Cherenkov radiation, under acceleration, while simultaneously taking into account recoil and how these inherently quantum mechanical processes alter the emission spectrum.	

\section{Cherenkov emission via accelerated currents}

Building upon the formalism of the previous section, let us now endeavor to compute the emission spectra by an accelerated electron current in the Cherenkov regime. The overall prescription will be to utilize the kinematics of Rindler space to determine trajectories, velocities, etc. and incorporate their dynamics into the emission. Recalling the response function, Eq. (\ref{response}), is given by
\bqa
\Gamma = q^{2} \int d\xi e^{-i\Delta E \xi}  U_{\m \nu}[x',x]G^{\m \nu}[x',x]. \label{response1}
\eqa
Note that we are now integrating over the proper time, also parametrized by $\xi$, of the electron. To analyze accelerated Cherenkov emission we integrate over the proper time rather than laboratory time since the accelerated trajectories are parametrized this way. We will begin by noting the velocity tensor in now given by $U_{\m \nu} = u_{\m} u_{\nu}$. Recalling that under proper acceleration $\bar{a}$, the four-velocities at proper time $\tau$ will be given $u^{\m} = (\cosh{(\bar{a} \tau)},0,0,\sinh{(\bar{a} \tau)})$ \cite{winitzki}. Computation of the emission spectrum, parametrized by the lab frame acceleration $a$, with $\bar{a}=a\gamma^3$, then yields the result,
\bqa
\frac{d\Gamma}{d\omega}&=& \frac{\alpha}{2 \beta \gamma^{2}}\lbk \lb 2\gamma^{2} - 1 \rb \lbk 1- \cos^{2}{(\theta_{c})} \rbk  - \sqrt{\lbk 1- \cos^{2}{\theta_{c}} + \lb \frac{a \gamma^{2}}{n\beta \omega } \rb^{2} \rbk^{2} +4 \cos{\theta_{c}} \lb \frac{a \gamma^{2}}{n\beta \omega } \rb  } \rbk. \label{aqrate}
\eqa
This emission spectrum is the main result of this work and illustrates the nature of accelerated-Cherenkov radiation. Note that it reduces to the standard Frank-Tamm formula, Eq. (\ref{ft}), as the acceleration goes to zero. We also recall the standard Cherenkov angle, as derived in the previous section, is given by $\cos{\theta_{c}}= \frac{1}{n\beta}+\frac{\Delta E}{n\beta \omega \gamma}$. The presence of acceleration then yields the new acceleration-dependent Cherenkov angle,
\bqe
\cos{(\theta_{a})} = \sqrt{\cos{\theta_{c}}^2 + \lb \frac{a\gamma^{2}}{n\beta \omega}  \rb^{2}}.\label{acos}
\eqe
Note, not only have we successfully recovered the Frank-Tamm formula with an additional acceleration-dependent term but also have the added functionality of the Unruh-DeWitt and its ability to incorporate recoil \cite{ginsburg, ido}. Overall, we see the that presence of acceleration serves to not only change the emission spectrum but also the emission angle. These effects are not merely a byproduct of having a time-dependent velocity but exist independently of the standard Cherenkov effect entirely. One such effect is to produce a low frequency cutoff in the emission spectrum. When this effect is combined with the recoil correction, which produces a high energy cutoff, we will find a resonance or local maximum in the spectrum. Figure \ref{emission} below details the general structure of Cherenkov radiation with contributions from both acceleration and recoil. Note that both contributions are required to produce a local maximum. Here, and throughout the manuscript, we assume a constant index of refraction. 
\begin{figure}[H]
\centering  
\includegraphics[scale=.39]{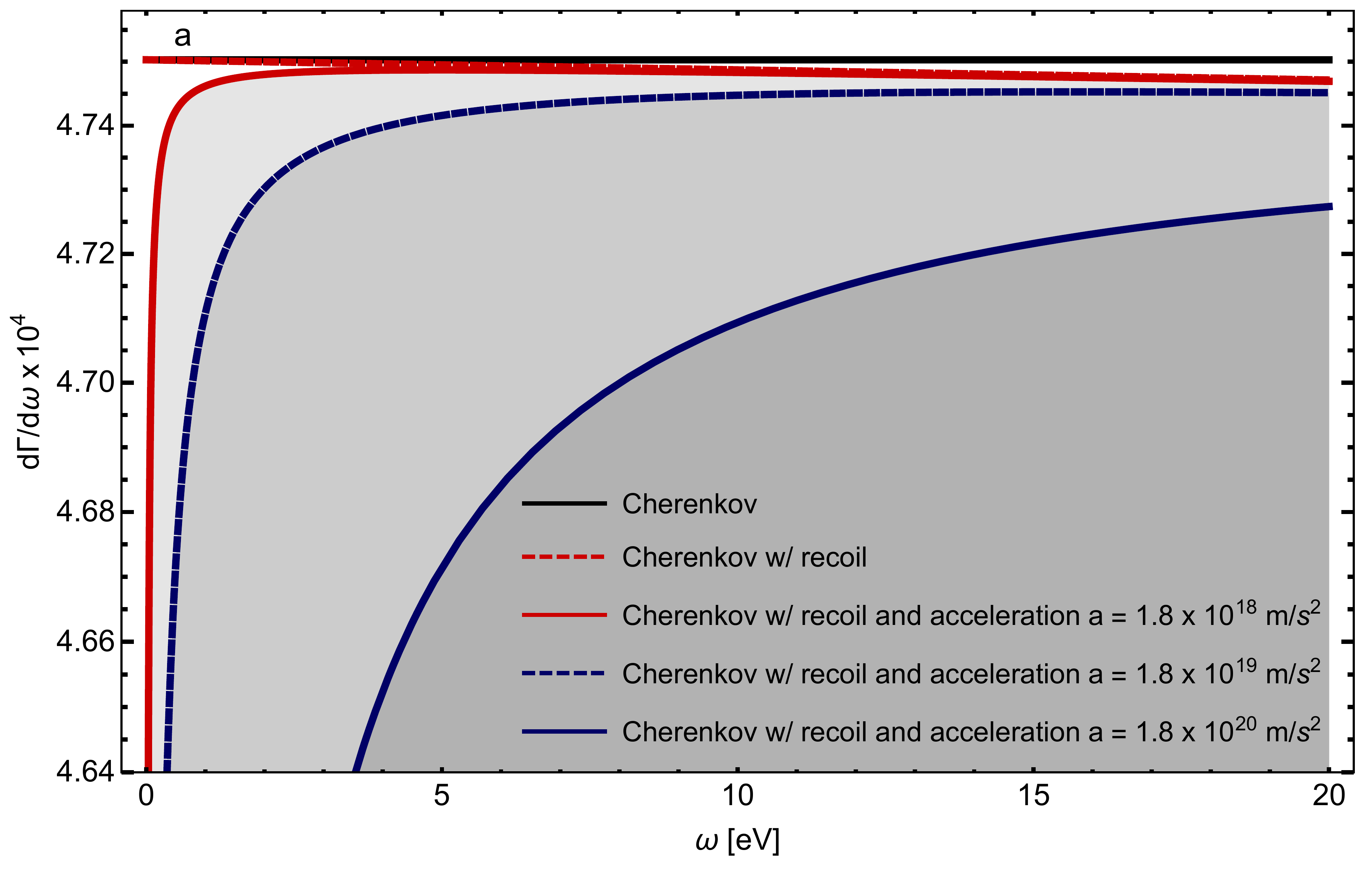}
\end{figure}
\begin{figure}[H]
\centering  
\includegraphics[scale=.39]{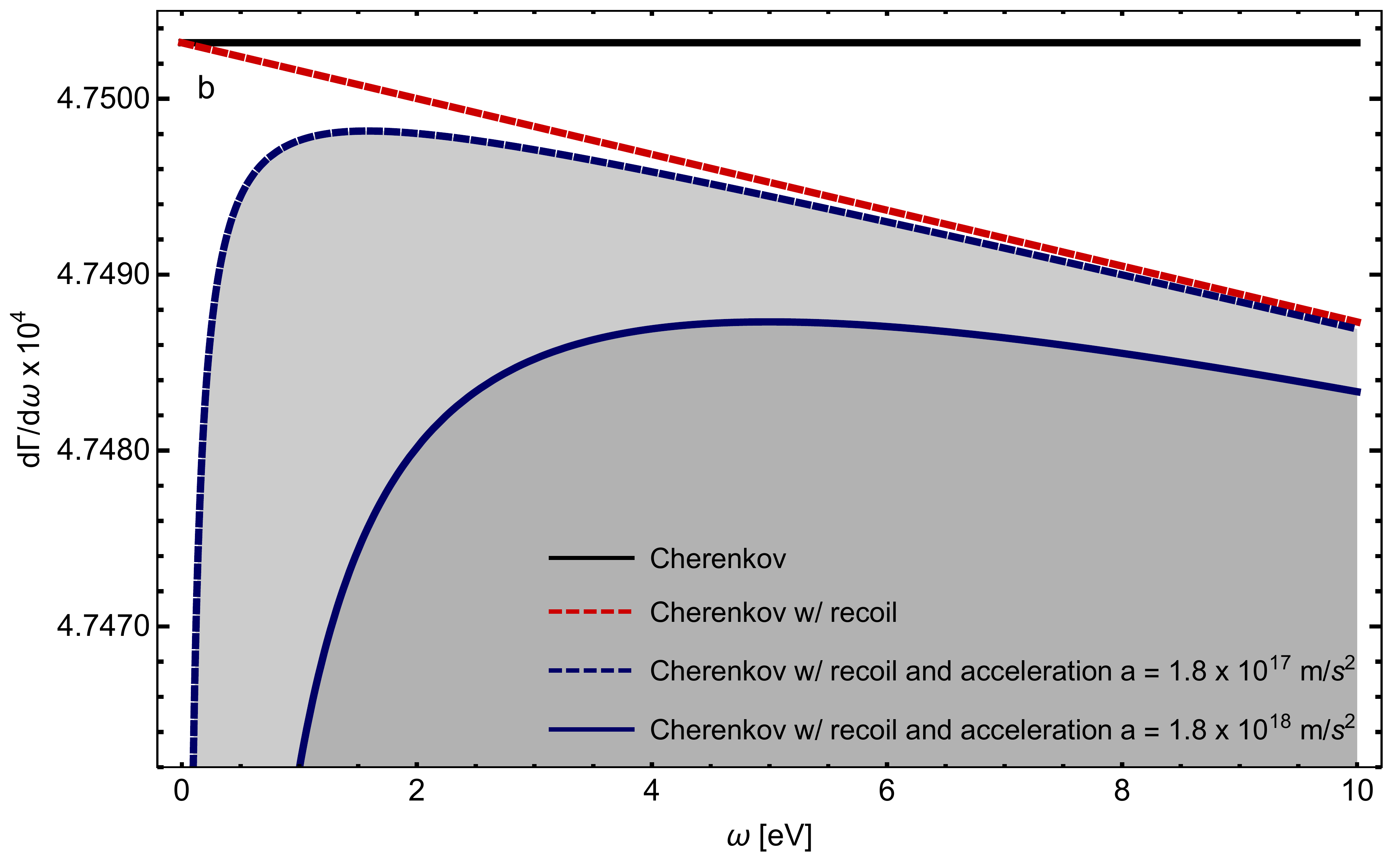}
\caption{The structure of Cherenkov spectra with and without acceleration and recoil. We assume a constant index $n = 1.5$, $\beta = .7$, and with an array of experimentally realizable accelerations. In (a) we see the overall large scale structure and in (b) we have zoomed in on the resonance peak. The acceleration scale $1.8 \times 10^{18}$ m/s$^{2}$ corresponds to a laser field amplitude of 1 GeV/m, an acceleration energy scale of $\omega_{a} = 5.5 \times 10^{-4}$ eV, and a cutoff frequency, see Eq. (\ref{cutoff}), of $\omega_{c} = 1.7 \times 10^{-3}$ eV. Note that both acceleration and recoil are required to see the peak in the emission frequency.}	
\label{emission}
\end{figure}
Without acceleration this resonance structure vanishes. We see that the recoil correction serves to give the Cherenkov spectrum a monotonically decreasing dependence on the frequency. Consequently, a tilted spectra is, modulo other effects such as material dispersion, a signature of recoil in Cherenkov emission. However, with acceleration present in the system we will see that there is clearly a local maximum in the spectrum which depends explicitly on the acceleration. This local maximum requires both acceleration and recoil to be present in the system and therefore may provide a new scenario to explore the phenomena of radiation reaction. To better characterize this process let us examine the peak and cutoff structure. The subsequent expressions, neglect the last term in the square root of  Eq. \ref{aqrate}, i.e. the cross term, so as to examine analytical forms of the dominant contributions in both the low and high frequency cutoff limits. As such, for the low energy cutoff, i.e. when the emission rate goes to zero, we find the following condition
\bqa\label{cutoffcos}
\omega_{c}  =  \frac{a\gamma}{\sqrt{2 }n \beta^{2} }\frac{1}{\sqrt{1-\cos^{2}{(\theta_{c})}}}.
\eqa
Without an internal energy level, i.e. no anomalous Doppler effect, we have the following cutoff 
\bqa
\label{cutoff}
\omega_{c}  &=& \omega_{a} \frac{1}{\sqrt{(n\beta)^2 -1  }}.
\eqa
Here we defined the characteristic energy scale set by the particle's acceleration, $\omega_{a} = \frac{a\gamma}{\sqrt{2 } \beta}$. Note the presence of a resonance at the Cherenkov condition.  Moreover, the overall scale of the cutoff frequency from Eq. (\ref{cutoff}) is specific to having an accelerated particle, e.g. an electron, with characteristic energy $\omega_{a}$ in the Cherenkov regime. As with the Unruh effect \cite{unruh1}, the energy scale $\omega_{a}$ set by acceleration is often vanishingly small. The important point however, is that we have found that Cherenkov emission provides another arena to explore acceleration energy scales which may be more experimentally tenable than in the case of the Unruh effect. As an example, the resonance at the superluminal threshold can potentially be used to make this acceleration scale more accessible. It is important to note that an acceleration energy scale of $\omega_{a} = 5.5\times 10^{-4}$ eV used as the characteristic energy scale in our figures is based off an acceleration of, $a \sim 1.8 \times 10^{20}$ m/s$^{2}$, that is indeed experimentally realizable in common laser systems with an amplitude of $\sim 1$ GV/m. Moreover, table top plasma wakefield accelerators are even capable of producing accelerations, $a \sim 10\times 10^{22}$ m/s$^{2}$ \cite{rosenz}. These systems may also provide a setting to explore this cutoff structure of Cherenkov emission. Unlike the above case with no internal energy levels, i.e. with $\Delta E = 0$, the inclusion of an internal energy gap we will yield a quartic equation that determines the cutoff frequency in the presence of radiation reaction and acceleration. As before, the inclusion of recoil is accomplished by setting the energy gap to $\Delta E = \frac{\omega^{2}(n^{2} - 1)}{2m}$ in the Cherenkov angle. The fact that we will have a quartic cutoff is due to the frequency dependence in the recoil correction. As such, the cutoff frequency is determined by the solution to the following equation,
\bqa\label{ca}
\omega_{c}  &=& \frac{\omega_{a}}{\sqrt{\lb n\beta \rb^2 - \lb 1 +\frac{\omega_{c}(n^{2}-1)}{2m\gamma } \rb^2}} 
\eqa
Moreover, the presence of recoil also produces a high frequency cutoff even in the absence of acceleration. From Eq. (\ref{aqrate}) we can determine the high frequency cutoff due to recoil (in agreement with \cite{ido}) to be,
\bqa\label{cr}
\omega_{c'}  &=&\frac{ 2m\gamma (n\beta - 1)}{n^2-1}.
\eqa
We can also compute the peak frequency that is produced by the simultaneous presence of acceleration and recoil. A local maxima in  the Cherenkov spectrum in these systems therefore implies both acceleration and recoil. The peak itself is determined by computing the maximum of our generalized Frank-Tamm formula Eq. (\ref{aqrate}). It resultant expression is quartic in nature, $\frac{\omega_{a}^{2}}{\omega^{3}} = \frac{(n^2-1)}{2 \gamma m} \lbk 1 + \frac{\omega(n^{2}-1)}{2 m\gamma}  \rbk$,  but has an approximate solution given by the following equation,
\bqe
\omega_{p} = \lbk \frac{2 m \gamma \omega_{a}^{2}}{n^{2}-1}  \rbk^{1/3}.
\label{resonance}
\eqe
The photons of this peak frequency will be the only photons emitted at the Cherenkov threshold. Consequently, by modulating the electrons velocity and acceleration at or near the threshold, one can essentially tune the Cherenkov radiation to a unique frequency that is also emitted into a unique direction, i.e. controlling the electrons velocity can put the system at resonance and controlling the acceleration can tune the peak frequency emitted by setting the energy scale $\omega_{a}$. As such, with a sufficiently intense source of modulated electrons one could create a tunable Cherenkov emitter. In Fig. \ref{plot2} we examine the structure of the cutoff frequency produced by both the recoil and the acceleration as an electrons velocity approaches Cherenkov threshold from above for an index of refraction of $n = 1.5$. The peak resonance that is converged to is given by the solution of Eq. (\ref{resonance}) and gives the peak frequency emitted for any superluminal velocity under the assumption of a constant index. As such, we note that by examining the solution to the cubic resonance equation, we can tune the dominant emission frequency by driving both the velocity and acceleration of a Cherenkov system at the superluminal threshold. Moreover, we note that the index of refraction will also depend on frequency. This can also be taken into account by insertion of the frequency dependent index of refraction into the cutoff and peak frequency equation and numerically solving that equation as well. Any vertical slice of Fig. \ref{plot2} will yield the maximum frequency as well as the high and low frequency cutoffs. The full cutoff curve is determined by Eq. (\ref{ca}). The asymptotic low frequency cutoff produced by the acceleration is determined by Eq. (\ref{cutoff}). The asymptotic high frequency cutoff produced by the recoil is determined by Eq. (\ref{cr}). The maximum frequency emitted is determined by Eq. (\ref{resonance}) and it should be noted that this maximum determines the unique frequency that is emitted at the $\beta$ threshold and can be used to ``tune" the emission. This threshold is also shifted due to the presence of both acceleration and recoil. In addition, we can see from Fig. \ref{plot2} that radiation emission is suppressed at higher energies due to the recoil. This further serves to ensure that the electron, modeled as an Unruh-DeWitt detector, remains pointlike with a wavelength much smaller than that of the emitted photons. If this were not the case then we would have to take into account smearing of the detector and its modification on the emission spectrum; see e.g. \cite{eduardo} for a QED-based analysis. We should also comment that additional factors such as a strong dependence of the index of refraction on frequency or higher order processes could further modify the spectrum which would undoubtedly give rise to another rich area of investigation. 
\begin{figure}[H]
\centering  
\includegraphics[scale=.49]{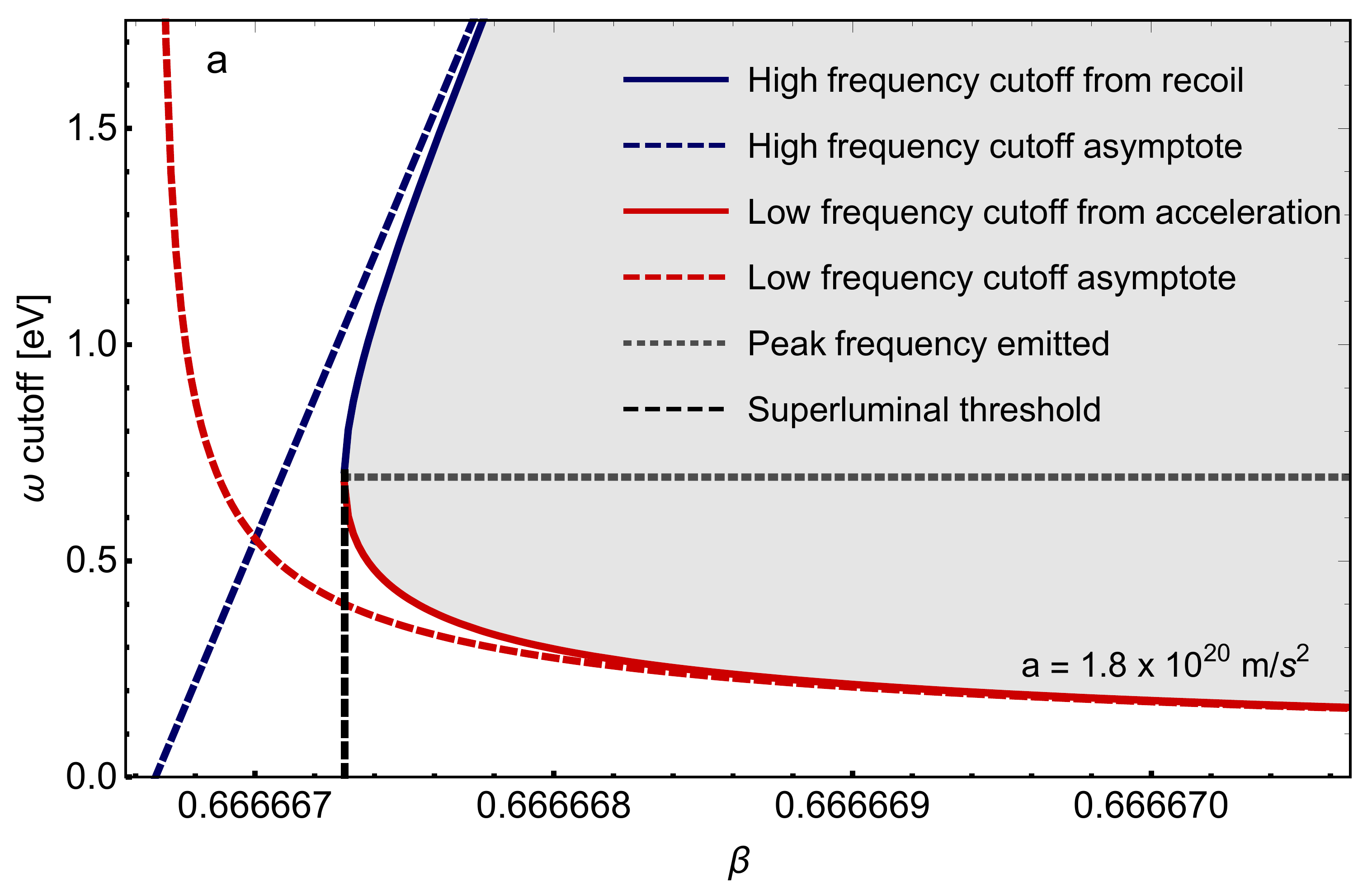}
\end{figure}
\begin{figure}[H]
\centering  
\includegraphics[scale=.49]{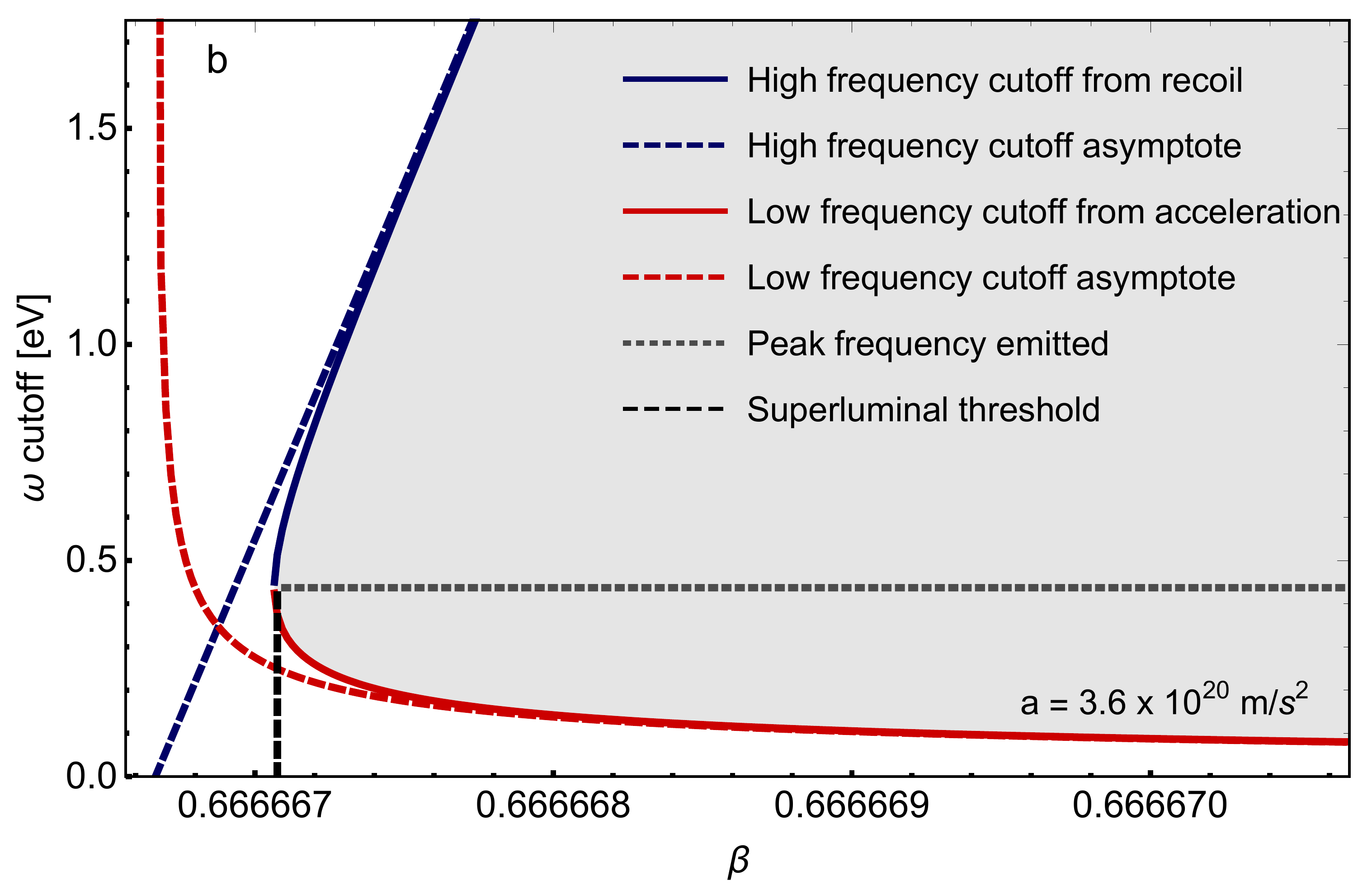}
\caption{The structure of Cherenkov spectral cutoffs and peak frequency emitted as a function of $\beta$ near the Cherenkov threshold. The shaded region determines the allowed frequencies of light emitted. The higher frequencies are bounded by the recoil correction while the lower frequencies are bounded by the acceleration energy scale $\omega_{a}$. These bounds converge to the peak frequency at threshold. The asymptotes for the low and high frequency cutoffs are given by Eqs. (\ref{cutoff}) and (\ref{cr}), respectively. The peak frequency is given by Eq. (\ref{resonance}) and the solid curve (both blue and red) of the spectral cutoff is given by Eq. (\ref{ca}). Here, we use the experimentally realizable parameters of $\omega_{a} = 5.5\times 10^{-4}$ eV (a), $\omega_{a} = 11\times 10^{-4}$ eV (b), and index of refraction $n=1.5$.} 	
\label{plot2}
\end{figure}
The ability to tune an electron's velocity with high precision at or near the Cherenkov threshold would imply the ability to control the peak frequency emitted. It is this peak frequency that the low and high energy cutoffs converge to at the threshold limit and will thus determine the only frequency emitted at resonance. The broadness of this peak will be determined by the control of the electron velocity; a sharp velocity distribution in an electron beam will then yield a sharp Cherenkov emission spectrum. As we drive the electrons with specific velocity and acceleration profiles we will get different peak frequencies. Figure \ref{plot3} shows the various peak frequencies as a function of velocity and acceleration at resonance; We substitute an index of refraction $n = 1/\beta$ everywhere. Note that for initial electron velocities and accelerations that are currently accessible in electron microscopes we would expect to see a peak frequency in, or near, the lower end of the visible spectrum. To be clear, this analysis assumes a relatively constant index of refraction as a function of frequency in the vicinity of the threshold. More specifically, given a beam velocity $\beta$ we must then select a material with index of refraction that will place the beam at or near the threshold, i.e. $n \sim 1/\beta$. Then, we must also require that material has an index that is fairly constant with respect to frequency, near the peak, in the vicinity of the threshold. With these conditions met, it stands to reason that a signal of the peak frequency can be measured. We can see from Fig. 4 that for typical electron velocities in electron microscopes $\beta \sim 0.6-0.7$ and with laser field amplitudes currently in regular use $\sim 1$ GV/m, we should then see a signal between $.1-1$ eV. This signal is well within the measurement capabilities of electron microscopes, e.g. an infrared detector or indirectly via electron energy loss spectroscopy \cite{abajo}. Moreover, these considerations can further add to the robustness of the Cherenkov effect's applicability to not only particle identification \cite{ginis, ido2} but also to radiation emission.  
\begin{figure}[H]
\centering  
\includegraphics[scale=.49]{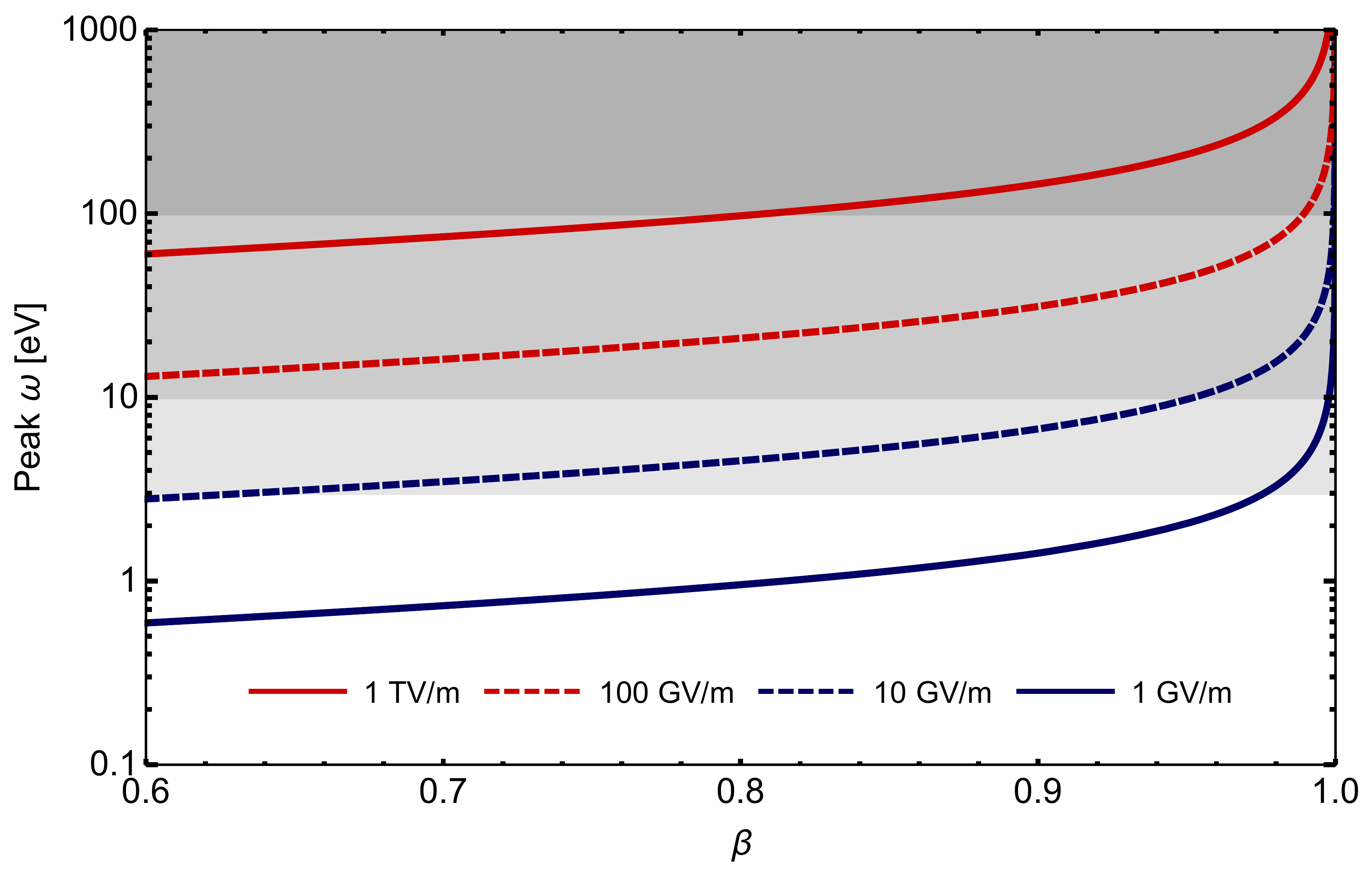}
\caption{The structure of the peak frequency as a function of the particle velocity and local acceleration. Here, the index of refraction is tuned to $n \sim \frac{1}{\beta}$ so we are always near resonance. Note that all velocity and acceleration profiles used are currently accessible experimentally and produce a peak frequency that is also well within current detection capabilities.} 	
\label{plot3}
\end{figure}
For the sake of completeness, we also consider the anomalous Doppler effect \cite{ginsburg, xihang}, i.e. an energy gap that does not depend on the emitted photons' frequency, we then have $\cos{\theta_{c}} = \frac{1}{n\beta} +\frac{\Delta E}{n\beta \omega \gamma} $ and a frequency cutoff given by following roots of Eqn. (\ref{cutoffcos});
\bqa
\omega_{c} &=& \frac{\frac{\Delta E}{\gamma} \pm \sqrt{ \lb \frac{\Delta E}{\gamma} \rb^{2} +\lb (n\beta)^2-1\rb \lb \lb \frac{\Delta E}{\gamma} \rb^2 + \omega_{a}^2 \rb}}{n^2\beta^2-1}.
\eqa
Again, we have a resonance but we see that the cutoff frequency can shift by any internal energy level. We also note the resonances can provide an additional method to probe the anomalous Doppler effect. Note that for both excitations and decays we potentially have two cutoff frequencies. This phenomenon exists both with and without acceleration. In the limit of zero energy gap we arrive back at the Cherenkov cutoff in the presence of acceleration, Eq. (\ref{cutoff}). The fact we have an acceleration energy scale $\omega_{a}$ is also reminiscent of the Unruh effect \cite{unruh1}. There is indeed a close relationship between the anomalous Doppler effect and the Unruh effect \cite{frolov1, frolov2} that has been explored in detail. The results presented here further stress the similarities between these two processes and outline the details of a new effect due to acceleration that is present in Cherenkov radiation. 

Finally, in addition to the novel structure of the Cherenkov emission spectrum in the presence of acceleration, there is also a deformation of the Cherenkov cone. If we are to assume an index of refraction that is independent of frequency, then the Cherenkov radiation will all be emitted into the well defined cone; the angle relative to direction of prorogation is defined by $\cos{(\theta_{c})} = \frac{1}{n \beta}$. When acceleration is present, it allows the Cherenkov cone to \textit{fill in} with the lower frequency end of the emission spectrum; see Fig. (\ref{cone}). Therefore, for a sufficiently thin sample, an accelerating charged particle in the superluminal regime, would reveal a filled-in circle rather than the standard ring in an imaging sensor.
\begin{figure}[H]
\centering  
\includegraphics[scale=.49]{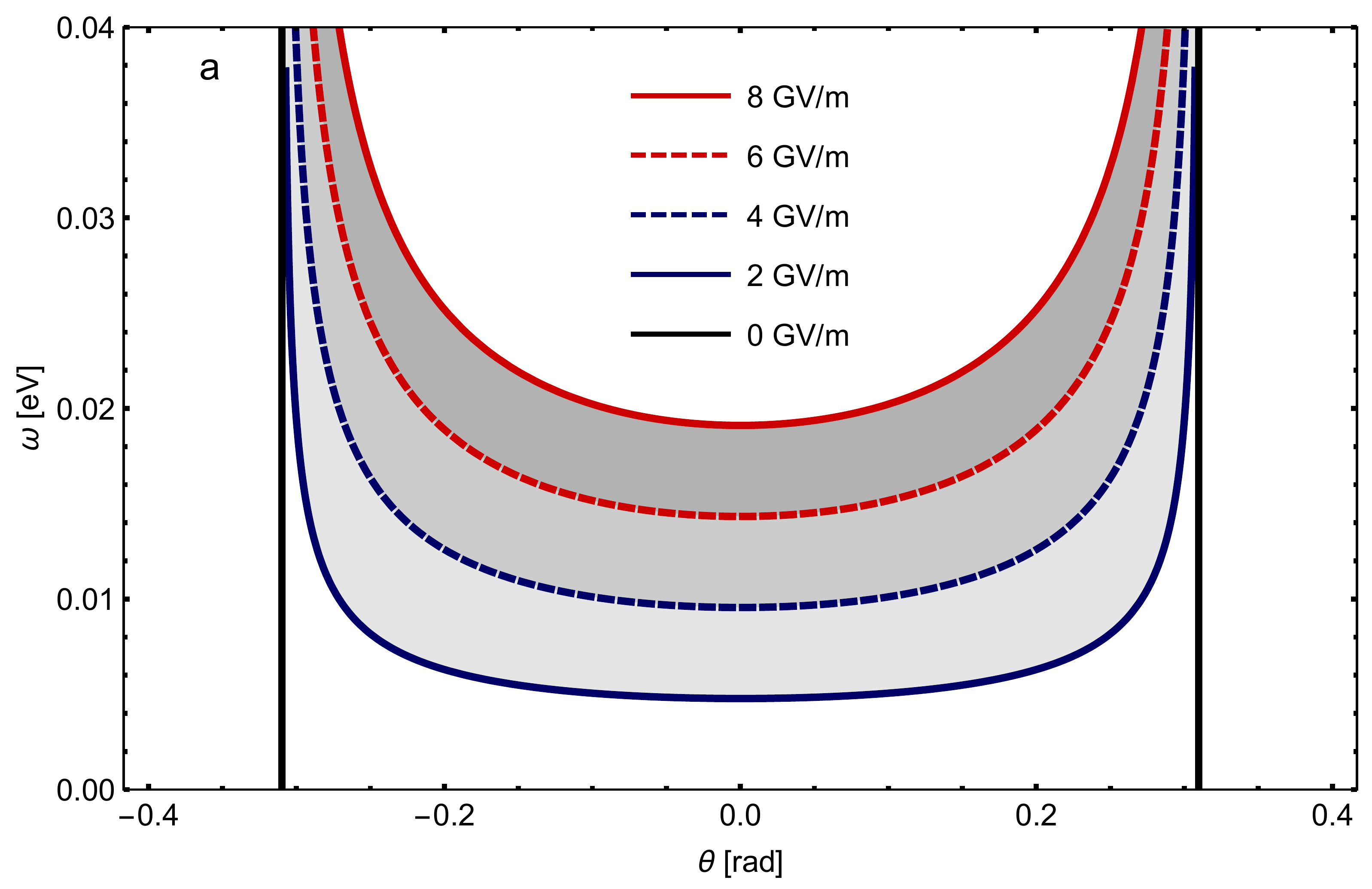}
\end{figure}
\begin{figure}[H]
\centering  
\includegraphics[scale=.49]{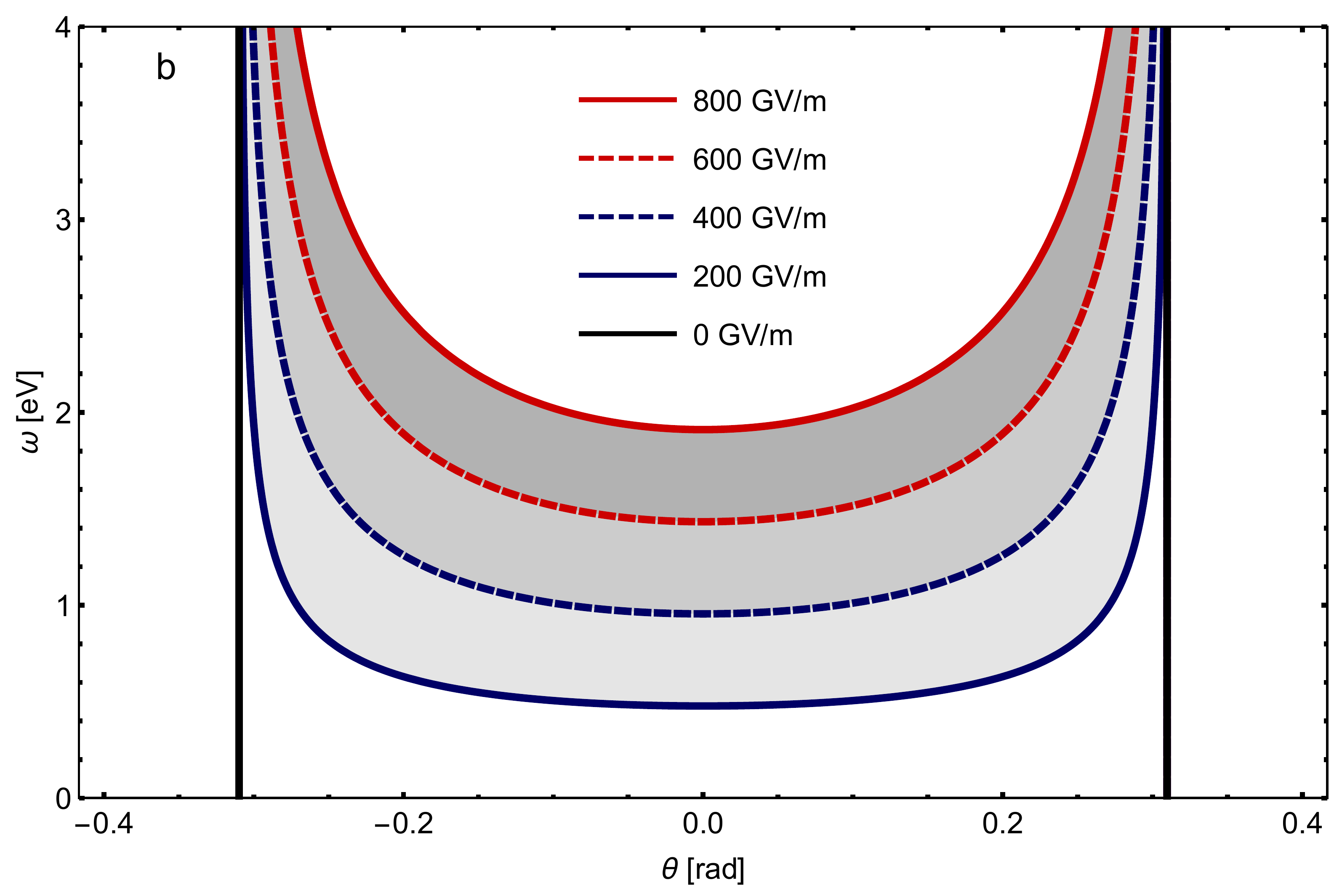}
\caption{The frequency-angle structure of the Cherenkov cone as a function of acceleration for a constant index of $n = 1.5$ and with $\beta = .7$, see Eq. (\ref{acos}). Note the effect of acceleration will be to fill in the Cherenkov cone with the lower energy end of the emission spectrum and thus produce a ``Cherenkov circle" rather than the standard Cherenkov ring.}	
\label{cone}
\end{figure}
To conclude this section, we have revealed a new aspect in the theory of Cherenkov emission which incorporates both acceleration as well as recoil. Its high tunability, accompanied by resonance peaks, may be employed as a high precision test of radiation reaction using Cherenkov emission under acceleration since the recoil is easily incorporated. We explored the various ways in which to characterize the radiation reaction, namely the acceleration-dependent cutoff frequency and local maxima in the spectrum. It is interesting to note the fact that we have an energy scale set by the acceleration is thus reminiscent of the Unruh effect \cite{matsas4}. 

Further implications on the effect of radiation reaction from charge renormalization would be interesting to investigate as well. Our work showed that the recoil correction, which is consistent with mass renormalization, manifests itself as a change in the electron's energy due to the recoil momentum imparted on the electron by the photon emission. The specific effect of charge renormalization \cite{peskin} would not only change the effective charge of the electron but should, in principle, contribute a new term in the energy gap as well. The charge functions as the monopole moment operator in the interaction and any renormalization of the charge may have an explicit presence there. Another potential correction, which is not discussed in this work, is field renormalization. That could provide another promising avenue to investigate sources of radiation reaction. In the end, we have that the utility of the Unruh-DeWitt detector applied to the problem of radiation reaction may be of particular value.

\section{conclusions}

In this manuscript we have made use of an Unruh-DeWitt detector to examine the effect of recoil and acceleration on Cherenkov radiation. Overall, our investigation has determined the effect of acceleration on both the emission spectrum as well as the emission angle. The combined presence of recoil and acceleration was then used to gain insight into the paradoxical nature of radiation reaction. We found a resonant energy cutoff resulting from the acceleration and tilted spectra due to the recoil which conspire together to produce a local maximum in the emission spectra. We also found that the presence of acceleration serves to fill in the Cherenkov cone with the lower frequency end of the emission spectrum. Having carried out the analysis via the use of an Unruh-DeWitt detector, we hope this work may open the way to expand its utility and also help develop a better understanding for the phenomenon of radiation reaction \cite{crystal}.

\section*{Acknowledgments}
We would like to thank Daniel Abutbul and Yarden Sheffer for many stimulating discussions. M.L. was supported in part at the Technion by a Zuckerman fellowship. E.C. was supported by the Canada Research Chairs (CRC) Program.

\goodbreak

\section{Appendix}
\beginsupplement
\setcounter{equation}{0}
\setcounter{figure}{0}
\setcounter{table}{0}
\setcounter{page}{1}
\makeatletter
\renewcommand{\theequation}{A\arabic{equation}}
\renewcommand{\thefigure}{A\arabic{figure}}

\subsection{Cherenkov Emission via Unruh-DeWitt detectors}

The emission of radiation produced by inertial currents offers the easiest setup to analyze recoil since the kinematics are purely Minkowskian. This setting provides a rather simple introduction to both radiation emission as well as the recoil. We begin by examining the emission of a photon in a refractive medium, $\hat{A}^{\m}(x)$, using the current interaction for QED \cite{peskin},

\bqe
\hat{S}_{I} = \int d^{4}x \hat{j}_{\m}(x)\hat{A}^{\m}(x).
\eqe

To model the electron current semi-classically we will make use of the weak recoil limit. To see how this is accomplished we begin by recalling that for spinors $u(s,p)$ and $v(s,p)$ of spin $s$ and momentum $p$ that are created by $\hat{a}^{\dagger}_{s,p}$ and $\hat{b}^{\dagger}_{s,p}$, we have the following electron field operators

\bqa
\hat{\psi}(x,t) &=& \int d^3 p \sum_{s}\lbk \hat{a}_{s,p}u(s,p)\phi_{p}(x,t) + \hat{b}^{\dagger}_{s,p} v(s,p)\chi_{p}(x,t) \rbk , \non \\
\hat{\bar{\psi}}(x,t) &=& \int d^3 p \sum_{s}\lbk \hat{a}^{\dagger}_{s,p}\bar{u}(s,p)\phi^{\ast}_{p}(x,t) + \hat{b}_{s,p} \bar{v}(s,p)\chi^{\ast}_{p}(x,t) \rbk.
\eqa

The positive and negative frequency modes are given by $\phi_{p}(x,t)$ and $\chi_{p}(x,t)$ respectively. Normally these modes are plane waves, however in more general spacetimes our only requirement is that they are positive and negative frequency modes with respect to the particle's/detector's proper time. Using these fields, we will formulate the electron current, $\hat{j}_{\m} = \hat{\bar{\psi}} \gamma_{\m} \hat{\psi}$. In order to enforce the weak recoil approximation we will ensure that our initial and final Fock states are labeled by their respective energies but have the same momentum, $p_{\m}$. The legitimacy of this approximation is restricted to the regime where the energy of the emitted radiation is much less than the rest mass of emitting the particle, e.g. $\sim 1 $ MeV for an electron. We will also neglect any spin effects which are described elsewhere, see e.g. \cite{ido}. Focusing strictly on electrons, i.e. no antiparticles, our current then reduces to the following

\bqa
\hat{j}_{\m}(x) = \bar{u}(p)\gamma_{\m}u(p)\phi^{\ast}_{E_{f}}(x,t)\hat{a}^{\dagger}_{p}\hat{a}_{p}\phi_{E_{i}}(x,t).
\eqa

Now, in keeping with the weak recoil approximation we make use of the Gordon identity \cite{peskin}. As such, we have $\bar{u}(p)\gamma_{\m}u(p) = u_{\m}$. We now make use of the fact that our positive and negative frequency mode solutions can be separated into their spatial and temporal components via $\phi(x,\tau) = g(x)e^{-iE\tau}$. We have chosen to parametrize our fields via the electron's proper time, $\tau$, to incorporate the Rindler coordinate chart when analyzing the accelerated case \cite{davies}. Our current now reduces to

\bqa
\hat{j}_{\m}(x) = u_{\m}g^{\ast}(x)g(x)e^{i E_{f} \tau}\hat{a}^{\dagger}_{p}\hat{a}_{p}e^{-i E_{i} \tau}.
\label{sdelta}
\eqa

We note that for sufficiently localized electronic wave functions, e.g. with a wavelength much smaller than the wavelength of emitted radiation, we have $g^{\ast}(x)g(x) = \delta{(x - x_{tr})}$ along the classical trajectory of the electron; which is assumed to be uniform in the weak recoil limit. This can also be seen by examining a smeared Unruh-DeWitt detector, see e.g. \cite{eduardo}. There, the smearing function which encodes a detector's spatial extent is given by $F(x) = -i\psi_{f}(x)\nabla \psi_{i}(x)$. Applying this formula for the inertial plane waves that are typically used to model electrons in the Cherenkov effect yields $F(x) = p g^{\ast}_{f}(x)g_{i}(x)e^{i\Delta E \tau}$, where $p$ is the momentum of the electron and $\Delta E = E_{f} - E_{i}$. Then, by inspection of Eq. (\ref{sdelta}), we see that our use of a delta function to model the semi-classical trajectory is equivalent to a detector with a sharply defined spatial extent. It would be interesting indeed to look for finite size effects, i.e. smearing, of an electron in the emission spectrum in processes like Cherenkov radiation. Finally, by attaching the time-dependence to the creation and annihilation operators we have

\bqa
 e^{iE_{f}\tau}\hat{a}^{\dagger}_{p}\hat{a}_{p}e^{-iE_{i}\tau} &=& e^{i\hat{H}\tau}\hat{q}(0)e^{-i\hat{H}\tau} = \hat{q}(\tau).
\label{smono}
\eqa

Here we defined the Heisenberg evolved charge monopole moment operator $\hat{q} = e^{i\hat{H}\tau}\hat{q}(0)e^{-i\hat{H} \tau}$ where $\hat{q}(0)$ is defined as $\hat{q}(0)\ket{E_{i}} = \ket{E_{f}}$ with $E_{i}$ and $E_{f}$ the initial energy and final energy of the electron moving along the trajectory, $x_{tr}$, of the current. The energy gap of the detector is $\Delta E = E_{f} -E_{i}$ and the charge $q$ of the electron is given by $q =\vert \bra{E_{f}} \hat{q}(0)\ket{E_{i}} \vert$. As such, we have transformed our fermionic current into a semi-classical charged current coupled to an Unruh-DeWitt detector \cite{davies, unruh1, matsas1, matsas2, matsas3}

\bqe
\hat{j}_{\m}(x) = u_{\m}\hat{q}(\tau)\delta^{3}(x-x_{tr}). 
\eqe

This formalism allows us to examine transitions both up and down in energy. In the case of uniform motion through an indexed medium, transitions up in energy are associated with the anomalous Doppler effect \cite{ginsburg}. This extra degree of freedom will allow us to incorporate electronic excitations to higher energy, i.e. the recoil. With the intent to examine Cherenkov radiation, i.e. photon emission from a charge moving faster than the speed of light in medium, we formulate the following amplitude;

\bqe
\mathcal{A} = i\bra{\mathbf{k}}\otimes \bra{E_{f}}\hat{S}_{I}\ket{E_{i}}\otimes \ket{0}.\label{sdiffprob}
\eqe

The square of the this quantity determines the transition probability per momentum of the final state photon. Note, by having a detector we are relieved of integrating over the final electron momentum states. The final momentum is essentially fixed by the detector gap and is tuned directly to the energy change experienced during the recoil. This is equivalent to a delta function density of states, $ \frac{d\mathcal{P}}{d^{3}k} = \int d^{3}p_{f} \delta^{3}{(\mathbf{p}_{f} - \mathbf{p}_{r})} \vert \mathcal{A} \vert^{2} = \vert \mathcal{A} \vert^{2}.$ The nature of the recoil momentum $p_{r}$ will be explored later on when we define the Unruh-DeWitt detector energy gap. Using the mode decomposition for a massless vector field in a dielectric medium with index of refraction $n$, we have

\bqa
\hat{A}^{\m}(x) &=& \int \frac{d^{3}k}{(2 \pi)^{3/2}} \frac{\sum_{i}\epsilon_{i}^{\m}}{\sqrt{n^{2}2\omega}} \lbk \hat{a}_{k}e^{i(\mathbf{k}\cdot \mathbf{x} - \omega t) } + \hat{a}^{\dagger}_{k}e^{-i(\mathbf{k}\cdot \mathbf{x} - \omega t)}  \rbk \non \\
\eqa

Here we have a dispersion relation which relates the frequency to momenta via $n\omega = k$ \cite{jackson}. This non-trivial dispersion relation is at the heart of the recoil correction to radiation reaction and we will see how it leads to a nonzero recoil momentum by strict conservation of energy arguments; see section IV below. Constructing the probability from Eq. (\ref{sdiffprob}) yields

\bqa
\mathcal{P} &=& \int d^{3}k \Big|  \bra{\mathbf{k}}\otimes \bra{E_{f}}\int d^{4}x \hat{j}_{\m}(x)\hat{A}^{\m}(x)\ket{E_{i}}\otimes \ket{0} \Big|^{2} \non \\
&=& \int d^{3}k\int d^{4}x\int d^{4}x'\Big|  \bra{E_{f}} \hat{j}_{\m}(x)\ket{E_{i}} \vert^{2} \vert \bra{\mathbf{k}} \hat{A}^{\m}(x)\ket{0} \Big|^{2}.
\label{scurrent}
\eqa

Evaluation of the square of the electron current matrix element yields

\bqa
\Big|   \bra{E_{f}} \hat{j}_{\m}(x)\ket{E_{i}} \Big|^{2} &=& \Big|  \bra{E_{f}} u_{\m}(x)e^{i\hat{H}\tau}\hat{q}(0)e^{-i\hat{H} \tau}\delta^{3}(x-x(t))\ket{E_{i}} \Big|^{2} \non \\
&=& q^{2} u_{\m}(x)u_{\nu}(x') \delta^{3}(x-x_{tr}(t))\delta^{3}(x'-x'_{tr}(t')) e^{-i\Delta E(\tau'-\tau)} 
\eqa

Next, we shall evaluate the vector field inner product. For this we will need to integrate over the final state momenta, thereby developing the total emission probability. The result is the Wightman function \cite{davies} for the photon. Hence,

\bqa
\int d^{3}k  \vert \bra{\mathbf{k}} \hat{A}^{\m}(x)\ket{0} \vert^{2} = \bra{0} \hat{A}^{\dagger \nu}(x')\hat{A}^{\m}(x)\ket{0}.
\eqa

Here we have factored out the completeness relation, $\int dk \ket{k}\bra{k} = 1$ over momentum eigenstates. The resultant two point function, with vector indices, factors into a scalar two point function weighted by the sum over indexed vector polarizations. Evaluation of this two point function, and being sure to keep polarization sum inside the integral, and without evaluating the integral over the momentum yields

\bqa
\bra{0} \hat{A}^{\dagger \nu}(x')\hat{A}^{\m}(x)\ket{0} &=& \bra{0} \int \frac{d^{3}k'}{(2 \pi)^{3/2}}\frac{\sum_{i}\epsilon_{i}^{'\nu}}{\sqrt{n'^{2}2\omega'}}\lbk \hat{a}_{k'}e^{i(\mathbf{k'}\cdot \mathbf{x'} - \omega' t') } + \hat{a}^{\dagger}_{k'}e^{-i(\mathbf{k'}\cdot \mathbf{x'} - \omega' t')}  \rbk \non \\
&\;\;\;\;\; \times& \int \frac{d^{3}k}{(2 \pi)^{3/2}} \frac{\sum_{i}\epsilon_{i}^{\m}}{\sqrt{n^{2}2\omega}} \lbk \hat{a}_{k}e^{i(\mathbf{k}\cdot \mathbf{x} - \omega t) } + \hat{a}^{\dagger}_{k}e^{-i(\mathbf{k}\cdot \mathbf{x} - \omega t)}  \rbk \ket{0} \non \\
&=& \frac{1}{(2 \pi)^{3}}\frac{1}{2} \int \frac{d^{3}k}{n(\omega) \omega}\sum_{ij}\epsilon_{i}^{\m}\epsilon_{j}^{\nu '} e^{i(\mathbf{k}\cdot \Delta \mathbf{x} - \omega (t'-t))}.
\eqa

Combining all the pieces we can formulate the total transition probability. Hence,

\bqa
\mathcal{P} & =& \int d^{4}x\int d^{4}x'\Big|   \bra{E_{f}} \hat{j}_{\m}(x)\ket{E_{i}} \Big|^{2} \Big| \bra{\mathbf{k}} \hat{A}^{\m}(x)\ket{0} \Big|^{2} \non \\
& =& q^{2}  \frac{1}{(2 \pi)^{3}}\frac{1}{2}\int d^{4}x\int d^{4}x'  \int \frac{d^{3}k}{n(\omega)\omega}u_{\m}(x)u_{\nu}(x') \sum_{ij}\epsilon_{i}^{\m}\epsilon_{j}^{\nu '}  \non \\
&\;\;\;\;\; \times& \delta^{3}(x-x_{tr}(t))\delta^{3}(x'-x_{tr}'(t'))  e^{-i\Delta E  (\tau'-\tau)+i(\mathbf{k}\cdot \Delta \mathbf{x} - \omega (t'-t))} \non \\
& =& q^{2}  \frac{1}{(2 \pi)^{3}}\frac{1}{2}\int dt dt' \int \frac{d^{3}k}{n(\omega)\omega} \frac{u_{\m}(x)u_{\nu}(x')\sum_{ij}\epsilon_{i}^{\m}\epsilon_{j}^{\nu '}}{u_{0}(x)u_{0}(x')}  e^{-i(\Delta E(\tau '-\tau)- \mathbf{k}\cdot \Delta \mathbf{x}_{tr} + \omega (t'-t))}.
\label{sprob}
\eqa

Note, in the last line we changed our integration variable from proper time to coordinate time via $dt = u_{0} d\tau$. Then, by defining the difference and average laboratory times, namely $\xi = t' - t$ and $\eta = (t' + t)/2$ respectively, we can decouple the temporal integrations \cite{lynch, lynch1}. Note, this parameterization has a unit Jacobian, i.e. $dt dt' = d\xi d\eta$. In the presence of trajectories that are linear in time, i.e. unaccelerated, the trajectory will obey the following relation $\Delta x = x(t') - x(t) = x(t' - t) = x(\xi)$. As a consequence we will have $\Delta x_{tr} = \Delta x_{tr}(\xi)$. Moreover, with a trajectory that is constant in velocity we will have $\epsilon_{i}^{\m}$ and $u_{\m}(x)$ independent of time but dependent on the photon emission angle. Noting the resultant expression is therefore independent of the laboratory time $\eta$ we can formulate the transition rate via $\frac{d\mathcal{P}}{d\eta} = \Gamma$. As such,

\bqe
\Gamma = q^{2}  \frac{1}{(2 \pi)^{3}}\frac{1}{2}\int d\xi \int \frac{d^{3}k}{\omega} \frac{u_{\m}(x)u_{\nu}(x') \sum_{ij}\epsilon_{i}^{\m}\epsilon_{j}^{\nu '}}{u_{0}(x)u_{0}(x')} e^{-i(\frac{\Delta E}{\gamma} \xi - \mathbf{k}\cdot \Delta \mathbf{x}_{tr}(\xi) + \omega \xi)}.
\eqe

We have defined the $\xi$-dependent Lorentz gamma $\gamma$ in the last line to boost the proper time difference to the lab frame. We can then define the scalar quantity $U = \frac{u_{\m}(x)u_{\nu}(x')\sum_{i,j}\epsilon_{i}^{\m}\epsilon_{j}^{\nu}}{u_{0}(x)u_{0}(x')}$. Moving into momentum space spherical coordinates, with the $z$-axis aligned along the direction of the velocity, we have

\bqa
\Gamma &=&q^{2}  \frac{1}{(2 \pi)^{2}}\frac{1}{2}\int d\xi \int d\cos{\theta} d\omega U\omega n   e^{-i(\frac{\Delta E}{\gamma}\xi-\omega n \Delta x_{tr}(\xi)\cos{(\theta)} + \omega  \xi)}. 
\eqa    

Here we used the dispersion relation $\omega = k/n$. Now, for inertial spacetime trajectories under constant velocity with respect to the lab frame we have $\Delta x_{tr}(\xi) = \beta \xi $. As such we can integrate over time along the trajectory yielding

\bqa
\Gamma &=& q^{2}  \frac{1}{(2 \pi)^{2}}\frac{1}{2}\int d\xi \int d\cos{\theta} d\omega U\omega n  e^{-i(\frac{\Delta E}{\gamma}- \omega n \beta  \cos{(\theta)} + \omega  )\xi} \non \\
&=& q^{2}  \frac{1}{4 \pi}\int d\cos{\theta} d\omega  U\omega n(\omega)  \delta(\frac{\Delta E}{\gamma} - \omega n \beta\cos{(\theta)} + \omega  ).
\label{sdirac1}
\eqa

Let us examine the scalar quantity $U$ in more detail. Since it is the 4-velocity of the charged particle contracted with the polarization vector we can evaluate it using the properties of the polarization. Since the polarization vector has no zero component and will always be orthogonal to the wave vector of the photon, we find

\bqa
U &=& \lb \sum_{i} \frac{u_{\m}}{u_{0}}\epsilon^{\m}_{i} \rb \lb \sum_{j} \frac{u_{\nu}}{u_{0}}\epsilon^{\nu}_{j} \rb^{\dagger} = \beta^{2} \sin^{2}{(\theta)}.
\eqa

Here we see the $\sin{(\theta)}$ rather than $\cos{(\theta)}$ comes out of the dot product. This is because the angle $\theta$ lies between the charged particle velocity and momentum of the photon. Since the polarization is everywhere perpendicular to the momentum we get $\cos{(\pi/2-\theta)} = \sin{(\theta)}$ from the dot product. As such, our emission rate now becomes	

\bqa
\Gamma &=& q^{2}  \frac{\beta^{2}}{4\pi}\int d\cos{\theta} d\omega  \sin^{2}{(\theta)} \omega n\delta(\frac{\Delta E}{\gamma}- \omega n \beta  \cos{(\theta)} + \omega ).
\eqa

Prior to integrating over the angle we note the presence of the delta function which enforces conservation of energy contains the celebrated Cherenkov emission angle deformed by the presence of the energy gap, this is the anomalous Doppler effect \cite{ginsburg, xihang}. Hence, the root yields the Cherenkov angle,  

\bqe
\cos{\theta} = \frac{1}{n\beta}+\frac{\Delta E}{n\beta \omega \gamma}.
\label{scherenkov}
\eqe

If we are examining a recoil from a photon emission we can find an explicit form for the energy gap, see section IV below for further discussion. Considering an initial ``dressed" electron with total energy comprised of the electron rest mass and the emitted photons energy $E_{i} = \sqrt{m^2+\omega^2}$ and a final state electron energy comprised of a bare electron and recoil momentum $k=n\omega$ (where n is the frequency-dependent index of refraction) produced by the photon emission, $E_{f} = \sqrt{m^2+(n\omega)^2}$, see Fig. \ref{photon} for an illustrative example. The recoil momentum, $n\omega$, is characteristic of the non-trivial dispersion relation of the photon. As such, we can determine the energy difference to be

\bqa
\Delta E  \approx   \frac{(n^{2}-1)\omega^{2}}{2m}.
\eqa

From this relation we also obtain the fully relativistic quantum corrected Cherenkov emission angle as computed by Ginzburg \cite{ginsburg}. We also note this is consistent with mass renormalization where the electron's energy contains losses on account of it propagating in a material with dispersion \cite{tsytovich}. These losses manifest themselves as a frequency dependent change in the electron's mass. Our Cherenkov angle, including the recoil correction, is then given by

\bqe
\cos{\theta} = \frac{1}{n\beta}+\frac{(n^2-1)\omega}{2m\gamma n\beta}.
\label{sdoppler}
\eqe

Next, to integrate over the angles we note the root of the delta function, either given as above in the case of recoil or from Eq. (\ref{scherenkov}) for the more general anomalous Doppler effect. For angles which satisfy this root we define $\theta_{cr}$. We also pick up a factor of $\omega n \beta$ in the denominator from the Jacobian. As such we obtain

\bqa
\Gamma &=& q^{2}  \frac{\beta^{2}}{4\pi}\int d\cos{\theta} d\omega  \sin^{2}{(\theta)} \omega n \frac{\delta(\cos{(\theta)}-\cos{(\theta_{cr})} )}{\omega n\beta }  \non \\
&=& q^{2}  \frac{\beta}{4\pi}\int d\omega \lb 1- \cos^{2}{(\theta_{cr})} \rb .  
\eqa

We note the prefactor is nothing more than the fine structure constant $\alpha = \frac{q^{2}}{4 \pi}$. Moreover, we have now derived the Frank-Tamm formula generalized to the exact quantum result that includes the reaction of the particle to its recoil by the photon emission \cite{sokolov,ido}. Hence, 

\bqe
\frac{d \Gamma}{d \omega} = \alpha \beta  \lbk  1-\frac{1}{n^{2}\beta^{2}} \lb  1+\frac{(n^{2}-1)\omega}{m\gamma} + \frac{(n^{2}-1)^{2}\omega^{2}}{4m^2\gamma^{2}} \rb \rbk .
\eqe

To conclude this section, we have outlined the general procedure for the computation of the emission rate of Cherenkov radiation, including recoil, via the use of an Unruh-DeWitt detector. We outlined the general idea for how to incorporate recoil with the Unruh-DeWitt detector; this procedure is discussed in more detail in Section IV. What is most important is the fact that we are able to successfully incorporate recoil into the emission process. In the following section we will follow the same procedure while placing the current along an accelerated trajectory which is also superluminal. We then explore the emission spectrum of Cherenkov radiation, under acceleration, while simultaneously taking into account recoil and how these inherently quantum mechanical processes alter the emission spectrum.	

\subsection{Cherenkov emission via accelerated currents}

Building upon the formalism of the previous section, let us now endeavor to compute the emission spectra by an accelerated electron current in the Cherenkov regime. The overall prescription will be to utilize the kinematics of Rindler space to determine trajectories, velocities, etc. and incorporate their dynamics into the emission. Recalling the photon emission probability, Eq. (\ref{sprob})

\bqe
\mathcal{P} = q^{2}  \frac{1}{(2 \pi)^{3}}\frac{1}{2}\int d\tau d\tau' \int \frac{d^{3}k}{\omega} \sigma^{\m} \sigma^{\dagger \nu} U_{\m \nu}  e^{-i(\Delta E \Delta \tau - \mathbf{k}\cdot \Delta \mathbf{x}_{tr} + \omega \Delta t)}.
\eqe

Note that we are back to integrating over the proper time of the electron and thus reabsorbed our $u_{0}$ components. To analyze accelerated Cherenkov emission we integrate over the proper time rather than laboratory time since the accelerated trajectories are parametrized this way. We will begin by examining the polarization vectors that are contracted with our velocity tensor. Recalling that under proper acceleration $\bar{a}$, the four-velocities at proper time $\tau$ will be given $u^{\m} = (\cosh{(\bar{a} \tau)},0,0,\sinh{(\bar{a} \tau)})$ \cite{winitzki}. Hence,

\bqa
U &=& \lb \sum_{i} u_{\m} \epsilon^{\m}_{i} \rb \lb \sum_{j} u_{\nu}\epsilon^{\nu}_{j} \rb^{\dagger} = \sinh{(\bar{a}\tau)} \sinh{(\bar{a}\tau')} |\sin{(\theta)}|^{2}. 
\eqa

Note, we have made explicit the magnitude squared of the sin term since our roots in the forthcoming delta functions will be complex in nature. From here we utilize the same change of variables to the difference and average proper times, $\xi'$ and $\eta'$. Moreover we will make use of the hyperbolic double angle formulas to obtain 

\bqe
\sinh{(\bar{a}\tau)} \sinh{(\bar{a}\tau')}  = \frac{1}{2}\lbk 2\cosh^{2}{(\bar{a}\eta')} -1 - \cosh{(\bar{a}\xi')} \rbk.
\eqe

This analysis is simplified using the time-dependent formalism presented in \cite{lynch1}. Combining all the above pieces we can now formulate the emission rate, $\Gamma_{\eta'} = \frac{d\mathcal{P}}{d\eta'}$. Thus

\bqa
\Gamma_{\eta'} &=& q^{2}  \frac{1}{(2 \pi)^{3}}\frac{1}{4}\int d\xi' \int \frac{d^{3}k}{n^{2}(\omega)\omega} \lbk 2\cosh^{2}{(\bar{a}\eta')} -1 - \cosh{(\bar{a}\xi')} \rbk  |\sin{(\theta)}|^{2} e^{-i(\Delta E \xi' - \mathbf{k}\cdot \Delta \mathbf{x}_{tr} + \omega \Delta t)} .
\eqa

Note that with respect to the proper time $\eta'$, the Lorentz gamma and velocity are given by $\gamma = \cosh{(\bar{a}\eta')}$ and  $\beta = \tanh{(\bar{a}\eta')}$ respectively. The trajectory is determined by $\Delta \mathbf{x}_{tr} =  \beta \Delta t $. The time difference is given by $\Delta t = \frac{2}{\bar{a}}\sinh{(\bar{a}\xi'/2)}\gamma$ for uniform acceleration and we will make use of the short time approximation such that $\Delta t = \xi' \gamma$. Now, to examine the momentum integrations we move to spherical coordinates with the momentum aligned along the $z$-axis. This will yield our emission spectrum $\frac{d\Gamma_{\eta'}}{d\omega}$. Hence

\bqa
 \frac{d\Gamma_{\eta'}}{d\omega} &=& q^{2}  \frac{1}{(2 \pi)^{2}}\frac{1}{4}\int d\xi' d\cos{(\theta)}n\omega\lbk 2\gamma^{2} - 1 - \cosh{(\bar{a}\xi')} \rbk  |\sin{(\theta)}|^{2} e^{-i(\Delta E    -n\omega \cos{(\theta)} \beta \gamma + \omega \gamma )\xi'}.
\eqa

The integration over proper time can now be evaluated by the use of delta functions. Note, the first term in the square bracket is the standard Cherenkov term and the hyperbolic cosine encodes the acceleration into the emission. Conversion of the acceleration dependent term into exponents and integration yields, 

\bqa
\frac{d\Gamma_{\eta'}}{d\omega}&=& q^{2}  \frac{1}{(2 \pi)^{2}}\frac{1}{4}\int d\xi' d\cos{(\theta)}n\omega\lbk 2\gamma^{2} - 1 -\frac{1}{2}e^{\bar{a}\xi'} - \frac{1}{2}e^{-\bar{a}\xi'} \rbk  |\sin{(\theta)}|^{2} e^{-i(\Delta E    -n\omega \cos{(\theta)} \beta \gamma + \omega \gamma )\xi'} \non \\
&=& q^{2}  \frac{1}{(2 \pi)}\frac{1}{4}\int  d\cos{(\theta)} |\sin{(\theta)}|^{2}n\omega\Big[\lb 2\gamma^{2} - 1 \rb  \delta{(\Delta E    -n\omega \cos{(\theta)} \beta \gamma + \omega \gamma  )}  \non \\
& & \;\;\;\; - \frac{1}{2}\delta{(\Delta E    -n\omega \cos{(\theta)} \beta \gamma + \omega \gamma +i\bar{a} )}   -\frac{1}{2} \delta{(\Delta E    -n\omega \cos{(\theta)} \beta \gamma + \omega \gamma -i\bar{a} )} \Big].
\label{sdirac2}
\eqa

Converting the delta functions for the subsequent $\cos(\theta)$ integration yields a Jacobian of $n\beta \omega \gamma$ for each delta function as well for each of the following three roots;

\bqa
\cos{\theta_{c}} &=& \frac{1}{n\beta} +\frac{\Delta E}{n\beta \omega \gamma} \non \\
\cos{\theta_{\pm a}} &=& \cos{\theta_{c}}\pm \frac{i\bar{a}}{n\beta \omega \gamma}.
\eqa

Recalling that $\sin{(\theta)} = \sqrt{1-\cos^{2}{(\theta)}}$. Evaluation of the angular integral then yields

\bqa
\frac{d\Gamma_{\eta}}{d\omega} &=& \frac{\alpha}{2 \beta \gamma}\lbk \lb 2\gamma^{2} - 1 \rb \lbk 1- \cos^{2}{(\theta_{c})} \rbk  - \sqrt{1-\lb \cos{\theta_{c}} + \frac{i\bar{a}}{n\beta \omega \gamma}\rb^{2}}\sqrt{1-\lb \cos{\theta_{c}} - \frac{i\bar{a}}{n\beta \omega \gamma}\rb^{2}}   \rbk \non \\
 &=& \frac{\alpha}{2 \beta \gamma}\lbk \lb 2\gamma^{2} - 1 \rb \lbk 1- \cos^{2}{(\theta_{c})} \rbk  - \sqrt{\lbk 1- \cos^{2}{\theta_{c}} + \lb \frac{\bar{a}}{n\beta \omega \gamma} \rb^{2} \rbk^{2} +4 \cos{\theta_{c}} \lb \frac{\bar{a}}{n\beta \omega \gamma} \rb  } \rbk.
\eqa

Recalling that the emission rate is parametrized by the electron's proper time, we need to boost back into the lab frame via $\frac{d \Gamma_{\eta'}}{d\omega} = \gamma\frac{d \Gamma}{\omega}$. We then have

\bqa
\frac{d\Gamma}{d\omega}&=& \frac{\alpha}{2 \beta \gamma^{2}}\lbk \lb 2\gamma^{2} - 1 \rb \lbk 1- \cos^{2}{(\theta_{c})} \rbk  - \sqrt{\lbk 1- \cos^{2}{\theta_{c}} + \lb \frac{\bar{a}}{n\beta \omega \gamma} \rb^{2} \rbk^{2} +4 \cos{\theta_{c}} \lb \frac{\bar{a}}{n\beta \omega \gamma} \rb  } \rbk.
\eqa

Care must be taken here since we have the proper acceleration and we need to boost back into the lab frame to have a well formulated emission rate. Recalling the relationship between the laboratory acceleration, $a$, and proper acceleration is given by $\bar{a} = a \gamma^{3} $ we have

\bqa
\frac{d\Gamma}{d\omega}&=& \frac{\alpha}{2 \beta \gamma^{2}}\lbk \lb 2\gamma^{2} - 1 \rb \lbk 1- \cos^{2}{(\theta_{c})} \rbk  - \sqrt{\lbk 1- \cos^{2}{\theta_{c}} + \lb \frac{a \gamma^{2}}{n\beta \omega } \rb^{2} \rbk^{2} +4 \cos{\theta_{c}} \lb \frac{a \gamma^{2}}{n\beta \omega } \rb  } \rbk.
\eqa

Note, we have successfully recovered the Frank-Tamm formula but now with an additional acceleration-dependent term. The above expressions shows the acceleration dependence in Cherenkov emission and, with the energy gap, can include the standard case, anomalous Doppler effect, and the recoil correction \cite{ginsburg, ido}. Here, and throughout the manuscript, we assume a constant index of refraction.

\setcounter{equation}{0}
\setcounter{figure}{0}
\setcounter{table}{0}
\setcounter{page}{1}
\makeatletter
\renewcommand{\theequation}{S\arabic{equation}}
\renewcommand{\thefigure}{S\arabic{figure}}

\end{document}